# Resilience assessment and improvement for electric power transmission systems against typhoon disasters: A data-model hybrid driven approach

Rui Yang [1], Yang Li[1,*]

[1] Key Laboratory of Modern Power System Simulation and Control & Renewable Energy Technology, Ministry of Education (Northeast Electric Power University), Jilin, 132012, China

* Corresponding author. E-mail address: liyang@neepu.edu.cn

**ABSTRACT:** In response to the damage to electric power transmission systems caused by typhoon disasters in coastal areas, a planning-targeted resilience assessment framework that considers the impact of multiple factors is established to accurately find the weak links of the transmission system and improve the system resilience. Firstly, this paper constructs the attenuation model of the wind field and the comprehensive failure model of the system, in which the model drive to establish the cumulative failure model of the transmission system is adopted, and multiple data-driven schemes to correct the cumulative failure rate by relying on the feature factor information of the transmission system is adopted. At the same time, an analytic hierarchy process-weighted arithmetic averaging (AHP-WAA) method is introduced to select the optimal data-driven evaluation scheme. Secondly, this paper adopts the impact-increment-based state enumeration (IISE) method to establish resilience indicators for systems and corridors separately. On this basis, the optimal promotion strategy is selected according to the construction difficulty, resilience improvement ability, and cost analysis. Finally, simulations on the IEEE RTS-79 system have been carried out considering the influence of the real typhoon scenarios, micro-topographic and transmission corridor information factors. The results demonstrate that the hybrid-driven system resilience assessment and improvement method can assist planners in accurately judging the resilience level of the system against typhoon disasters and selecting the best resilience improvement strategy based on the cost-effectiveness ratio.

## 1. Introduction

### 1.1 Background

China is one of the countries with the highest incidence of typhoons in the world. The strong winds, rainstorms, and other disasters caused by typhoon disasters pose a huge threat to the power system in the areas along the route and thereby cause great economic losses to society. In 2016, typhoon "Meranti" landed in Xiamen, Fujian, causing a total of 45 substations, 2,837 lines, 60,301 stations, and 3,312 million customers to lose power [1]. In August 2019, super typhoon "Likima" landed on the southeastern coast of China, causing more than 4,000 lines failures in Zhejiang and other provinces and power outages for 67,695 million customers [2]. Under the influence of global warming, the frequency of typhoon disasters affecting the power system in coastal areas has been increasing in recent years. More recently, with the integration of renewable energy resources [3-5], power system operation resilience is facing more challenges. Therefore, it is a pressing issue to be solved how to identify the fragile position of the transmission corridor and improve the resilience of the system quickly and accurately.

**1.2 Literature review**

Regarding the issue above, scholars have discussed presently how to accurately assess and improve the resilience of transmission systems against extreme natural disaster conditions. Due to the complexity and interdisciplinary nature of the topic, different researches are carried out in their respective fields. This paper divides the related research into three parts. i) Construction of fault models. ii) Establishment of resilience indicators. iii) Selection of transmission system optimization strategies [6, 7], as shown in Fig. 1.

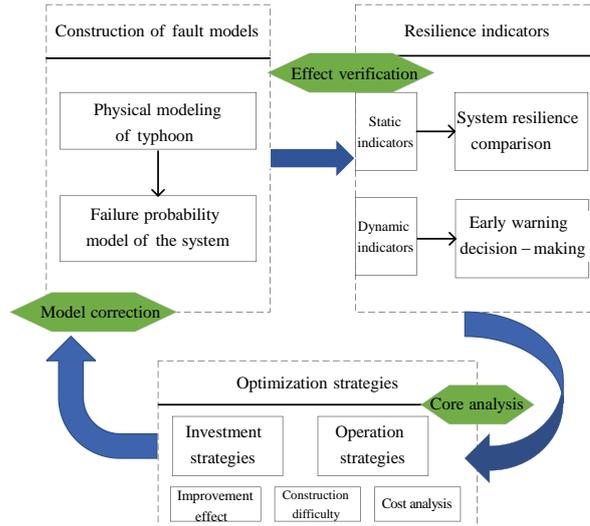

Fig. 1 Classification of researches conducted in the power system resilience

**1.2.1    Construction of fault model**

A typhoon is a composite meteorological disaster that can cause strong winds, rainstorms, floods, etc. The passing of typhoons will lead to a surge in the failure rate of transmission corridors. Therefore, it is necessary to establish a wind field model to analyze the wind speed in different regions, and then measure the failure rate of each transmission corridor. Batts in Ref. [8] proposed the first wind field model which adopted an empirical formula to simulate the change of wind speed. Ref. [9] on the bases of the Batts wind field model calculated the cumulative failure rate of transmission corridors by the integral theory and established a system resilience evaluation framework. However, the above studies only considered the impact of strong wind on the transmission system and ignored the ancillary disasters such as torrential rain, and the micro-topographic information of the transmission system. In Ref. [10], the authors fitted the relationship between the typhoon information, the micro-topographic information, and the failure rate of transmission corridors at different voltage levels based on the historical data of transmission corridors in coastal cities in Guangdong, and established a corresponding logistic regression model for the systematic risk assessment. Authors in Ref. [11]-[14] proposed a failure model of the transmission corridor according to the structural force theory, and Fu et al. in Ref. [11] adopted an evaluation method for tower collapse and disconnection accidents against typhoon and rainstorm disasters at the same time. By establishing the typhoon and rainstorm load force model, the dynamic response of transmission corridors against wind speed and rainstorm load were analyzed. In Ref. [12]-[14], the surrogate model was used to establish the fragile curve of the transmission corridor according to the simulated uncertain structural parameters. However, models that adopted structural force theory had high computational costs, so it is difficult to be used for large-scale studies. In addition, the model based on structural force theory could not capture some other information, such as micro-topographic, corridor information, etc. Authors in Ref. [15]-[17] simulated the failure rate of power components against typhoon disasters by Monte Carlo method based on a large number of historical typhoon key parameter data. In Ref. [18]-[20], the authors established a data-model-driven failure model of transmission corridors against typhoon disasters. However, the paper lacked the analysis of system resilience and corresponding improvement strategies.

**1.2.2    Establishment of resilience indicators**

The main content of power system resilience assessment is to establish a systematic resilience assessment index system and assessment theory. According to existing research, commonly used resilience evaluation indicators are usually divided into static indicators and dynamic indicators. Authors in Ref. [21, 22] adopted the Monte Carlo method to simulate various scenarios and calculated the static reliability index against different scenarios, thereby reflecting the system's overall resilience. However, as the reliability of the system improves, the sampling of abnormal states becomes more and more difficult, which greatly reduces the evaluation efficiency of the system. The state enumeration method is adopted by Ref. [23] to enumerate the dynamic reliability index of the system in various states, thereby reflecting the real-time flexibility of the system. This method is difficult to evaluate for large systems as the number of states grows exponentially with the components of the system. Authors in Ref. [24, 25] adopted the impact-increment-based state enumeration (IISE) to convert the traditional enumeration formula into a schema based on the impact -increments structure which simplified the calculation of state probability and greatly improve the evaluation efficiency and accuracy.

**1.2.3    Optimization strategies**

The existing optimization strategies are divided into investment strategies and operation strategies. Authors in Ref. [26]-[29] adopted an investment strategy to improve the resilience of the system by adding corridor redundancies and strengthening components. This method is expensive but effective. In Ref. [30]-[33], the authors adopted the method of optimizing the operation strategy, and introduced distributed generations in the fault system whose structure is changed to meet the needs of users to a certain extent but the access of the distributed generations increased the volatility of the power grid that made the resilience evaluation process more complicated [34].

Although there is much literature on the resilience analysis of transmission systems against typhoon disasters, there are still many problems in the existing research:

1) In terms of modeling, both the feature factors that affect the failure rate of transmission corridors and a comprehensive combination of the cumulative spatiotemporal impact of typhoons duration with the multi-factor information can create challenges when making the resilience evaluation model.

2) In terms of solution methods, it is difficult to seek the optimal evaluation results for different data-driven optimization schemes. In the above method research, Ref. [9] proposed a planning-targeted framework for assessing resilience and solved the spatiotemporal impact of wind speed on the resilience of the transmission system through a semi-empirical and semi-theoretical typhoon wind speed attenuation model. On this basis, this paper fully utilizes the historical data information of the transmission system and introduces a data-driven scheme in the traditional model-driven method as a correction to the failure rate of the transmission corridor. In addition, inspired by Ref. [19], the Gini index method, the OOB error method, and the entropy weight method are introduced to select the most accurate weight coefficients in the feature analysis. We can comprehensively find out the weaknesses of the system to improve the resilience of the system more accurately by reasonably establishing the failure model of the transmission system.

**1.3 Contributions of this paper**

The contributions of this paper are as follows:

1) This paper presents a data-model hybrid driven resilience assessment approach for electric power transmission systems by combining the optimal load reduction and comprehensive probabilistic failure model of the system. By exploring the relationship between the multi-factor information and the failure rate of transmission corridors, this approach can not only make full use of the model drive to model the physical laws of wind disasters and the mechanism of wind loads force, but also employ the data drive to analyze and construct predictive factors for predicting the failure rate of transmission corridors.

2) In view of the different judgment principles of various data-driven models, this paper proposes a multi-attribute decision-making method, namely analytic hierarchy process-weighted arithmetic averaging (AHP-WAA). This method adopts a combination of objective analysis of data weights and subjective human experience to seek optimal data-driven schemes.

3) Simulation tests on the IEEE RTS-79 test system attached on the coastline of Guangdong province, China have been performed. The results demonstrate the effectiveness and practicability of the proposed approach.

**1.4 Organization of this paper**

The rest of this paper is organized as follows: Section 2 mainly introduces the comprehensive failure probability model of the system based on the data-model hybrid drive, and Section 3 mainly studies the construction of system resilience evaluation indicators and improvement measures. Section 4 presents the IEEE RTS-79 test system case study, and Section 5 draws the conclusions.

**2. Construction of fault model**

The resilience of the transmission system refers to the ability of the transmission system to withstand, resist disturbances, and restore loads quickly [35]-[37]. Therefore, accurate measurement of the failure rate of the system against typhoon disasters is very important for the evaluation and improvement of the resilience of the system. The failure rate of the transmission system is not only related to the wind speed of the typhoon, but is also affected by the secondary disasters caused by the typhoon, the micro-topographic where the transmission corridor is located, and the transmission corridors information. The traditional model-driven approach is often difficult to consider multi-factor disaster scenarios simultaneously and faces problems such as difficulty in modeling weak-feature factors, while the complete data-driven method relies heavily on historical data. At present, the existing research is limited by the limited historical fault data, and it is difficult to evaluate the transmission system in a complete data-driven method. Therefore, this paper adopts a hybrid-driven method with the model drive as the main body and data drive as the correction which can more accurately simulate the impact of major meteorological disasters on the system and comprehensively consider the impact of other factors on the system, and improve the accuracy of resilience assessment to a certain extent.

**2.1 Model-driven failure model for transmission corridors**

Typhoon disasters are persistent disasters that cause cumulative damage to the power transmission system. Therefore, this section first builds a wind field attenuation model and simulates the influence of typhoon wind speed on the failure rate of transmission corridors based on the corridor stress model. Secondly, this section adopts the integral theory to analyze the cumulative failure rate of the transmission corridor during the typhoon.

**2.1.1 Wind field model**

The typhoon field model is a warm-core structure with low central pressure and high surrounding temperature. The formation of a wind field is usually determined by the historical data and the theoretical model on the premise of knowing the basic parameters of the typhoon. This paper selects the Batts wind field model that has more effective simulation results and high computing efficiency. The wind field model is shown in Fig. 2.

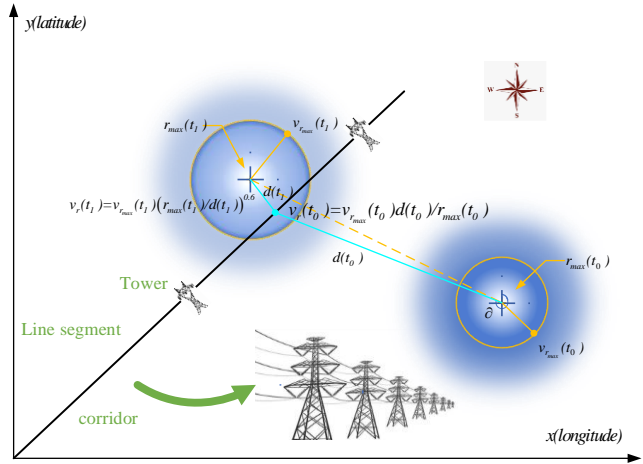

Fig. 2 Wind field model of the typhoon

In the model of the wind field, we assume that the pressure decays with time. The pressure decay model of the typhoon is given by

$$P(t) = \Delta P_0 - [0.02 + 0.02\sin(\partial)]t \qquad (1)$$

where $P(t)$ is the central pressure difference at moment $t$, $\Delta P_0$ is the initial pressure difference, and $\partial$ presents the angle between the direction of movement and the north direction.

According to the pressure decay model, the change of the maximum wind speed with time when the typhoon passes through is obtained by

$$v_{r_{max}}(t) = 0.865K\sqrt{\Delta H(t)} + 0.5v_T \quad (2)$$

where $K$ is the coefficient of the model, the value of $K$ is usually 6.93 at latitude 23ºN to 26ºN [4], and $v_T$ is the horizontal movement speed of the typhoon.

The real-time model of the maximum wind speed radius is obtained by [38]:

$$r_{max}(t) = \exp\left(2.63 - 5.086 \times 10^{-5} \Delta H(t)^2 + 0.0395 y(t)\right) \quad (3)$$

where $y(t)$ is the latitude of the typhoon center at moment $t$.

The real-time model of wind speed is [39]

$$v_r(t) = \begin{cases} v_{r_{max}}(t) d(t) / r_{max}(t), & d(t) \leq r_{max} \\ v_{r_{max}}(t)\left(r_{max}(t)/d(t)\right)^{0.6}, & d(t) > r_{max} \end{cases} \quad (4)$$

where $d(t)$ is the distance from the typhoon center to the transmission corridor at moment $t$, and $v_{r_{max}}(t)$ is the maximum wind speed of the transmission corridor at moment $t$.

### 2.1.2 Cumulative failure model of the transmission system

The damage to the transmission system is mainly reflected in the interruption of the line and the collapse of the tower. On the contrary, transformers and cables are hardly affected by typhoons. Therefore, this paper selects transmission towers and transmission lines as the analysis objects, and establishes the failure model of transmission corridors when the typhoon passes through. Because transmission corridors are usually long, different sections of the same corridor experience different wind speeds. Therefore, this paper treats the transmission corridor as a series combination of multiple tower-line analysis units. Each tower-line analysis unit includes a tower and a transmission line connected to it. For the convenience of analysis, this paper takes corridor $m$ as an example.

$$\lambda_{m,l}(t_i) = \exp\left[11 \times \frac{v_{m,l}(t_i)}{v_{d,line}} - 18\right] \Delta l \quad (5)$$

where $\lambda_{m,l}(t_i)$ is the failure rate of the $l-th$ section of corridor $m$ at moment $t$, $v_{m,l}(t_i)$ is the wind speed of the $l-th$ section of the corridor $m$, $v_{d,line}$ is the design wind speed of corridor $m$, and $\Delta l$ is the length of the $l-th$ section of corridor $m$.

The failure probability model of the $k-th$ transmission tower of the transmission corridor $m$ at moment $t_i$ based on the Batts typhoon model after the typhoon hit the coast.

$$\lambda_{m,k}(t_i) = \begin{cases} 0, & v_{m,k}(t_i) \in \left[0, v_{d,tower}\right] \\ e^{\gamma}\left[v_{m,k}(t_i) - 2v_{d,tower}\right], & v_{m,k}(t_i) \in \left[v_{d,tower}, 2v_{d,tower}\right] \\ 1, & v_{m,k}(t_i) \in \left[v_{ex}, \infty\right] \end{cases} \quad (6)$$

where $\gamma$ is the model coefficient, $v_{d,tower}$ is the design wind speed of the transmission tower.

In the failure probability model, the time is taken as the sequence, and the wind speed is the input quantity. The total continuous time is divided into multiple periods, and the cumulative failure rate of the transmission corridor is obtained.

From the beginning to the end of the typhoon landing, the cumulative failure rate of the $l-th$ section and the $k-th$ transmission tower of corridor $m$ are

$$p_{m,l} = 1-\exp\left(-\int_0^T \lambda_{m,l} dt\right)$$
$$p_{m,k} = 1-\exp\left\{-\int_0^T \left[\lambda_{m,k} / (1-\lambda_{m,k})\right]\right\} dt \quad (7)$$

Each corridor can be viewed as a series combination of multiple transmission towers and lines because each corridor tends to be far apart. When a certain section of the corridor fails, it means that the entire corridor fails. The failure probability model of the corridor is shown [40, 41]

$$p_m = 1 - \prod_1^K (1-p_{m,K}) \prod_1^T (1-p_{m,T}) \quad (8)$$

where $T$ is the number of line segments divided by corridor $m$, and $K$ is the number of transmission towers of corridor $m$.

**2.2 Data-driven failure model for transmission corridors**

This paper analyzes the correction coefficient for the area occupied by each tower-line analysis unit of corridor $m$, and forms the comprehensive failure rate of each tower-line analysis unit of corridor $m$, then combines the series formula of failure rate to obtain the comprehensive failure rate of corridor $m$.

First, we calculate the weights of feature factors to obtain three schemes by using Gini index method, OOB error method, and Entropy weight method. Second, we adopt the AHP-WAA method to select the optimal scheme based on experts' experience. Finally, the correction coefficient of the failure rate is obtained according to the tower-line information data of the actual system.

**2.2.1 Determination of feature factors**

The failure of the transmission corridor is usually determined by internal and external factors. The main external causes of line disconnection and tower collapse are the strong wind and the accompanying rainstorm, the environment around the transmission corridor, and other factors. The internal factors are mainly determined by the design wind speed, the operation time, and the type of transmission tower. Therefore, this paper takes the model-driven as the leading, establishes the wind load stress model of the transmission corridor, and adopts data-driven methods to obtain a correction coefficient of the failure rate of the transmission corridor to achieve scientific and accurate prediction. The feature factors are as follows:

Internal factors: design wind speed, operation time.
External factors: maximum wind speed, rainfall intensity, slope, wind angle, altitude.

Although the influential impacts of the above feature factors are different, they can be unified through machine learning. Fig. 3 depicts the relationship between the correction coefficient $k$ and feature factors of the transmission system failure probability based on the machine learning model [42].

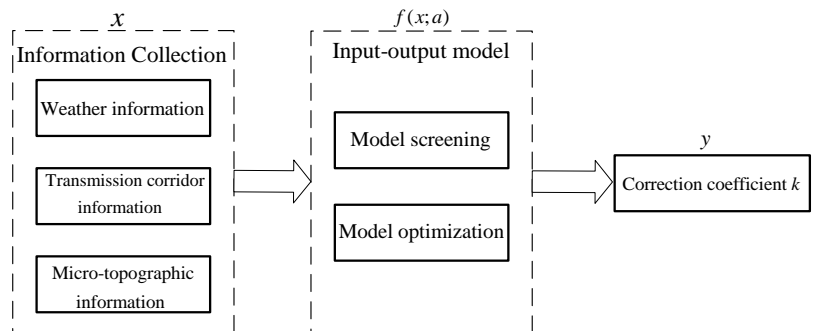

Fig. 3 The calculation process of the correlation coefficient

In Fig. (3), $x$ is the feature factor, $f(x;a)$ is the machine learning model, $y$ represents the correction coefficient $k$ of the failure rate.

### 2.2.2 Feature factors weight evaluation schemes

As is known, data-dirven methods have been successfully applied in power systems [43, 44]. Due to the different evaluation principles of different data-driven methods, the results of the evaluation are also different. In order to obtain the weight of the feature factor that is most in line with the objective reality, this paper selects the three most common data-driven methods in the current power system evaluation, namely the RF Gini index method, the RF OOB error method, and the entropy weight method. Based on the evaluation results of the three schemes and expert experience, the AHP-WAA is used to select the optimal driving scheme as the final decision scheme for the correction coefficient $k$.

Random Forest (RF) is an important ensemble learning method based on bagging, which is often used for classification, regression, and other problems. RF is composed of multiple binary trees. The root node contains all training samples. According to certain rules, each node selects the variable that minimizes the impurity of the branched node as the branch variable and continues to split until the stopping rule is met, and the specific process is shown in Fig. 4 [45, 46].

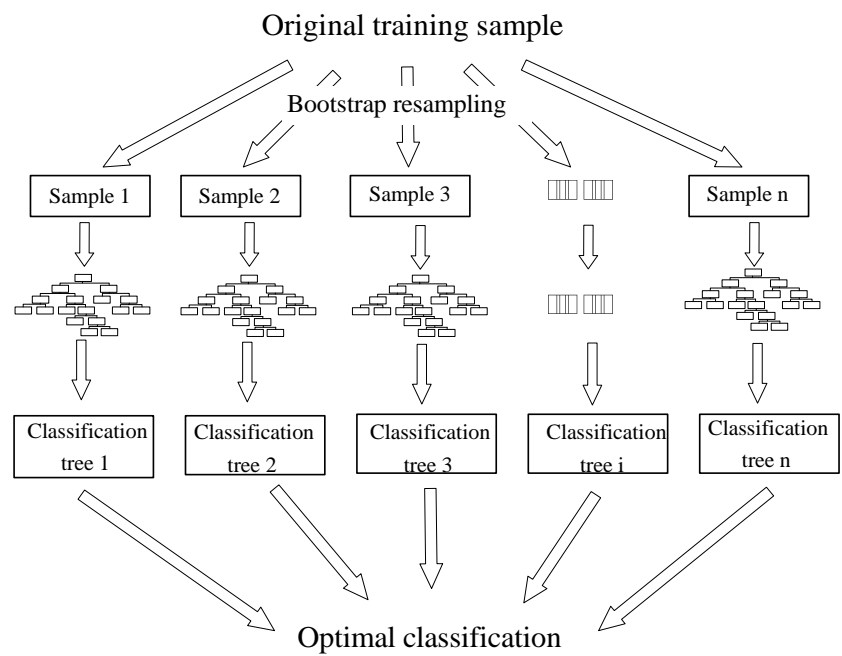

Fig. 4 Random Forest frame diagram

Because RF has the advantages of fast training speed, high processing dimension, high algorithm accuracy, and the ability to detect the influence between features during the training process, it is widely used in many fields such as finance, ecology, medicine, and cyberspace security. At present, the commonly used measures of impurity include Gini impurity and error rate.

This paper sets $m$ feature variables $y_1, y_2, \cdots y_m$, introduces the RF Gini method and OOB error method, and calculates the importance of the Gini method $VIM_j^{gini}$ and the importance of the OOB error method $VIM_j^{oob}$ for the feature variable $y_j$, respectively.

1) Gini index method

The principle of the Gini method: the average change of the node splitting impurity of the feature variable in the RF model [47]. The formula is as follows:

$$GI_m = 1 - \sum_{kI=1}^{|kI|} p_{mkI}^2 \qquad (9)$$

where $KI$ indicates that there are $KI$ categories, and the paper treats the transmission corridor fault situation as two categories (fault state $KI=0$, non-fault state $KI=1$), $P_{mKI}$ is the proportion of node m category $KI$.

Calculating the importance of feature variable $y_j$.

$$VIM_{jm}^{gini} = GI_m - GI_i - GI_r \qquad (10)$$

where $GI_i$ and $GI_r$ are the Gini values of the two nodes behind the branch.

As the RF contains $n$ decision trees, and the feature $y_j$ appears $R$ times in one of the decision trees, then its importance score in the RF model is

$$\begin{aligned} VIM_{ij}^{gini} &= \sum_{r=1}^{R} VIM_{jr}^{gini} \\ VIM_j^{gini} &= \frac{1}{n} \sum_{i=1}^{n} VIM_{ij}^{gini} \end{aligned} \qquad (11)$$

Normalizing the importance scores of all feature variables Gini index, that is, the weight of each feature.

$$VIM_j^{gini} = \frac{VIM_j^{gini}}{\sum_{i=1}^{m} VIM_i^{gini}} \qquad (12)$$

2) OOB error method

The principle of the OOB error method: the OOB error rate of the feature variable is affected by the noise introduced by the corresponding variable sample data. If the feature variable is important, the OOB error rate will change greatly after the noise is introduced.

When RF builds a binary tree, it uses a bootstrap sample to build a tree for the training set, and obtains the prediction OOB error rate. After adding noise to the sample of the feature variable $y_j$, RF builds the binary tree again, and calculates the corresponding prediction OOB error rate. The difference between the two OOB error rates is the permutation importance of the feature variable $y_j$ after normalization.

The importance of the feature $y_j$ in RF:

$$VIM_j^{OOB} = \frac{\sum_1^n (e_{j2} - e_{j1})}{n} \tag{13}$$

where $n$ is the number of decision trees, $e_{j1}$ is the OOB error rate of the feature variable $y_j$ in RF, and $e_{j2}$ is the OOB error rate after adding noise.

This method advances in high accuracy and simple algorithm but it is easily affected by data sets, and cannot get good results for the classification of small data sets and low-dimensional data sets.

3) Entropy weight method

The entropy weighting method is a popular weighting method at present, which fully utilize the sample information contained in each feature variable. Although it is based on objective data and is not affected by subjective factors, it ignores the importance of indicators, and it is difficult to reduce the dimension of evaluation indicators that often ignores the subjective wishes of decision-makers. In addition, the selected simple value is changed very little or suddenly larger and smaller, resulting in the determined indicator weights being to be far from the expected results [48, 49].

The main steps of the Entropy weight method are as follows:

Step 1: Normalize indicators. Since the measurement units of each indicator are not uniform, we need to standardize before calculating the comprehensive indicator.

$$x'_{ij} = \frac{\max\{x_{1j}, \ldots, x_{nj}\} - x_{ij}}{\max\{x_{1j}, \ldots, x_{nj}\} - \min\{x_{1j}, \ldots, x_{nj}\}} \tag{14}$$

where $j$ is the number of indicators, and $i$ is the number of samples of indicators.

Step 2: Calculate the entropy value of the $j-th$ index

$$e_j = -k \sum_{i=1}^n \frac{x_{ij}}{\sum_{i=1}^n x_{ij}} \ln\left(\frac{x_{ij}}{\sum_{i=1}^n x_{ij}}\right) \tag{15}$$

Step 3: Calculate the weight of each indicator

$$w_j = \frac{1 - e_j}{\sum_{j=1}^m 1 - e_j} \tag{16}$$

### 2.2.3 AHP-WAA multi-attribute scheme selection

The above-mentioned three methods are all objective weighting methods. Due to the different principles of the methods, the final weight evaluation results are different. This paper introduces the AHP-WAA scheme-making method to determine the optimal weight scheme. The AHP determines the relative importance of various factors in the hierarchy by pairwise comparisons [50]-[52]. In this paper, it is used to determine the score of feature variables in WAA multi-attribute scheme decision-making. This method is an analytical method that combines subjectivity and objectivity. When seeking the optimal scheme, both objective analysis and subjective experience evaluation are considered.

1) Constructing a decision matrix

This paper considers $n$ schemes for determining weights, $w$ feature variables, and constructs a decision matrix $Y$ according to the weight results determined by $n$ schemes.

$$Y = \begin{bmatrix} y_{11} & y_{12} & \cdots & y_{1w} \\ y_{21} & y_{22} & \cdots & y_{2w} \\ \vdots & \vdots & & \vdots \\ y_{n1} & y_{n2} & \cdots & y_{nw} \end{bmatrix} \quad (17)$$

where $y_{nw}$ is the weight value of the $w-th$ factor in the $n-th$ scheme.

2) Calculating scheme attribute weights

First, we set $\omega_i = \dfrac{\sum_{i=1}^{n} y_{iw}}{\sum_{i=1}^{n}\sum_{w=1}^{w} y_{iw}}$ to determine the average weight of each feature variable in the $n$ schemes and perform expert scoring analysis based on the average weights of feature variables and the experience of experts, and then construct a discriminant matrix of relative importance weights $A = (a_{ij})_{w \times w}$, where $a_{ij}$ is the ratio of the $i-th$ feature variable importance index to the $j-th$ index. Finally, the matrix $A$ is normalized and added by row to obtain the relative weight of the feature variable $q = [q_1 \ q_2 \ \cdots \ q_w]^T$.

3) Calculating the weight vector

$$D = Y \times q = [d_1 \ d_2 \ \cdots \ d_n]^T \quad (18)$$

We select the scheme according to the maximum value in a matrix $D$ as the final decision scheme.

**2.2.4　Calculation of correction coefficient**

According to the optimal evaluation scheme determined above, the correction coefficient $k$ of the failure rate is obtained from the weight corresponding to each factor. The steps are as follows:

1) Determine the boundary value of the comprehensive score

The effect of the numerical growth of different feature variables on the correction rate of the transmission corridor is different. The rainfall intensity, maximum wind speed, slope, wind angle, and altitude are positively correlated with the fault correction rate of the transmission corridor, while the operation time and design wind speed are negatively correlated. For this reason, when calculating the boundary value, the sign of the feature variable with positive correlation is '+', and the sign of the feature variable with negative correlation is '-'. At the same time, the value range of the feature variable based on the actual situation is selected as shown in Table 1, and the comprehensive score boundary value is calculated

$$W_{B_{\max}} = \max\left(\sum_{i=1}^{m} \omega_i x_{B_{\max},i}, \sum_{i=1}^{m} \omega_i x_{B_{\min},i}\right)$$
$$W_{B_{\min}} = \min\left(\sum_{i=1}^{m} \omega_i x_{B_{\max},i}, \sum_{i=1}^{m} \omega_i x_{B_{\min},i}\right) \quad (19)$$

where $\omega_i$ is the weight of the feature variable $i$, $m$ is the number of feature variables, and $x_{B_{\max},i}$, $x_{B_{\min},i}$ are the boundary quantity set by the feature variable $i$.

2) Calculate the composite score

$$W = \omega x^{\mathrm{T}} = \sum_{i=1}^{m} \omega_i x_i \tag{20}$$

3) Obtain the correction coefficient

According to reference [19, 59], this paper maps the value of the revision coefficient $k$ to the interval [0.9-1.4]

$$k = 0.5 \frac{W - W_{B_{\min}}}{W_{B_{\max}} - W_{B_{\min}}} + 0.9 \tag{21}$$

Table 1 Variable value range

| Maximum wind speed (m/s) | Design wind (m/s) | Rainfall intensity (mm/h) | Wind angle (°) | Altitude (m) | Operation time (year) | Slope (°) |
| --- | --- | --- | --- | --- | --- | --- |
| 0~60 | 20~50 | 0~60 | 0~180 | -20~150 | 0~40 | 0~180 |

## 2.3 Calculation of the comprehensive failure rate

Considering the reduction of computational cost and the insignificant change of feature factors within a certain range, this paper adopts the grid method to divide the analysis area into 1km×1km. In this section, we focus on the influence of wind speed on the transmission corridor when the typhoon passes, calculate the cumulative failure rate of the transmission corridor through a model-driven approach, and adopt the design wind speed, the operation time, the maximum wind speed, the rainfall intensity, the slope, the wind angle, and the altitude experienced by transmission corridor as feature factors, then calculates correction coefficient $k$ in a data-driven method. The comprehensive failure rate of corridor $m$ in the hybrid-driven model is

$$\begin{aligned} p_{mc,i} &= k_{m,i} p_{m,i} \quad i = 1, 2, \cdots, n \\ P_{mc} &= 1 - \prod_{i=1}^{n} (1 - p_{mc,i}) \end{aligned} \tag{22}$$

where $k_{m,i}$ is the correction coefficient of tower-line analysis unit $i$ in corridor $m$, $p_{m,i}$ is the cumulative failure rate of tower-line analysis unit $i$, $p_{m,i}$ is the comprehensive failure rate of the tower-line analysis unit $i$, $n$ is the number of tower-line analysis units of corridor $m$.

Fig. 5 depicts the construction process of the comprehensive failure probability model for the corridor $m$.

Step 1: The typhoon wind field attenuation model is constructed based on the Batts model by collecting the typhoon disaster information.

Step 2: The research area is divided by the grid method, and the model-driven method is used to analyze the influence of wind speed on the failure rate of the transmission corridor.

Step 3: In order to measure the continuous impact on the transmission corridor when the typhoon transits. In this paper, the integral theory is used to obtain the cumulative failure rate of the corridor.

Step 4: According to the collected historical data, the training set of feature factors that affects the failure rate of the transmission system is determined.

Step 5: A variety of data-driven schemes are used to determine the weight of feature factors. This paper adopts the AHP-WAA to select the weight evaluation results of the optimal data-driven scheme.

Step 6: The feature factor weight is combined with the specific tower-line analysis unit information of corridor $m$ to determine correction coefficient $k_{m,i}$ of tower-line analysis unit $i$, and comprehensive failure rate $p_{mc,i}$ of unit $i$ in corridor $m$ is calculated.

Step 7: All tower-line analysis units of corridor $m$ are corrected to obtain comprehensive failure rate $p_{mc}$ of corridor $m$.

In the same way, we respectively traverse the correction coefficients of tower-line analysis units in other corridors by obtaining the information data of other corridors, and then the comprehensive failure probability model of the system is constructed.

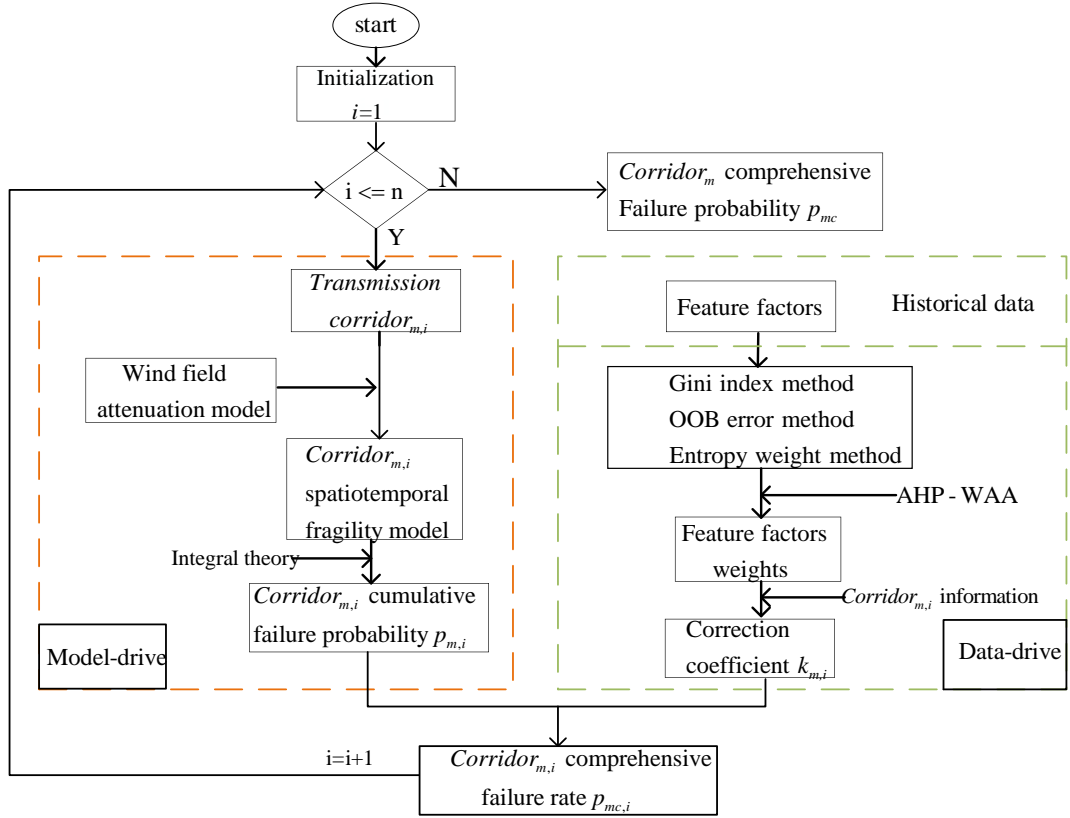

Fig. 4 Comprehensive failure probability model for corridor $m$

## 3. Resilience assessment and improvement methods

The establishment of resilience assessment indicators helps quantify system resilience and finds its weaknesses. From the view of the system point, we can measure the amount of system performance decline caused by typhoons within the planning expectations through system indicators. From the perspective of the corridor, we can identify the weak corridors of the system through corridor-indicator, so as to improve the resilience of the system through a targeted optimization strategy.

### 3.1 Analysis of typhoon probability model

The research adopts the state enumeration method to analyze the typhoon occurrence probability. According to the empirical distribution in Ref. [39], the moving speed $v_T$ and the central pressure difference $\Delta H_{0,w}$ show a log-normal distribution relationship, and the moving direction of the typhoon $\theta$ shows a bi-normal distribution. The empirical distributions of other versions do not affect the model framework proposed in this paper. Taking the wind speed of a typhoon as an example, with $c_H$ as the moving step, the integral theory is shown

$$P_r(\Delta H_{0,w}) = \int_{\Delta H_{0,w}-c_H/2}^{\Delta H_{0,w}+c_H/2} f(\Delta H_0) d\Delta H_0 \tag{23}$$

where $\Delta H_{0,w}$ is the central pressure difference, $P_r$ is the probability of occurrence of $\Delta H_{0,w}$.

Similarly, we can calculate the probability of occurrence of the $v_T$ and $\theta$. The above key parameters are divided into several equal parts, we choose a group from each key parameter, and combine them into a typhoon probability model. The probability of key parameters of typhoon $w$ occurring in a specific area is shown in Fig. 6. The probabilistic model is as follows:

$$P_w = P_r(\Delta H_{0,w}) P_r(v_{T,w}) P_r(\theta_w) \tag{24}$$

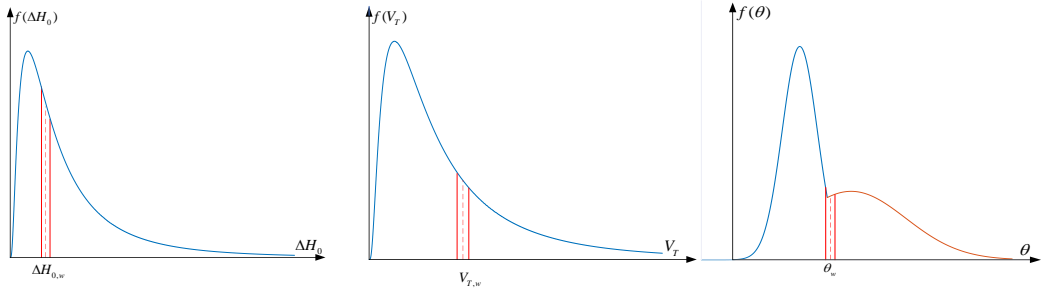

Fig. 6 Occurrence probability of key parameters of typhoon $w$

### 3.2 The system-level resilience indicator

System-level resilience evaluation metrics are crucial for quantifying the transmission system resilience. As shown in Fig. 7, now the resilience triangle is widely used in the existing literature as a method to quantify system resilience. The principle adopts the integral of the performance reduction and time as the system resilience index. The larger the index, the lower the resilience of the system [55].

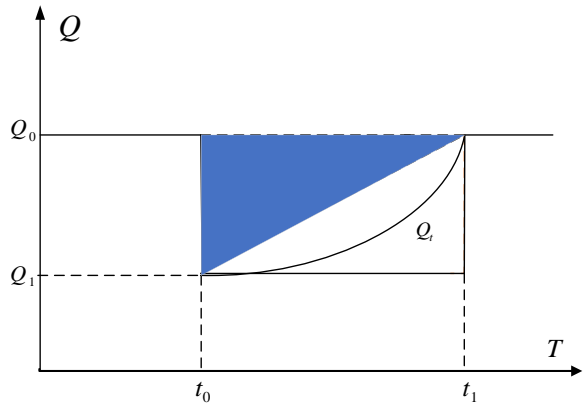

Fig. 7 Resilient triangle and rectangle

When the typhoon passes, $Q_0$ is the performance state of the system during normal operation, $Q_1$ is the state when the system performance is the most serious during the typhoon, $Q_t$ is the performance curve. Its resilience index calculation formula can be expressed

$$\Upsilon = \int_{t_0}^{t_1} [Q_0 - Q(t)] dt \tag{25}$$

Noticeably, the uncertainty of typhoon for the above indicators are not considered. Generally, typhoon disasters are different, and the unified quantification is inaccurate. Decision-makers often consider the most severe performance degradation when considering the performance of transmission systems to ensure that the transmission equipment is as safe as possible. In Fig.7, this study changes the resilience triangle into the resilience rectangle. The revised resilience index calculation formula is simplified as follows:

$$R = E[\Upsilon] \approx E[Q_0 - Q_1] = \sum_{w \in W} P_w (Q_0 - Q_1)_w \qquad (26)$$

where $E$ is the expected value of the minimum load reduction of the system, $P_w$ is the probability of the occurrence of typhoon $w$, and $W$ is the set of potential typhoons.

This article adopts the IISE to solve $E[\Upsilon]$. The method greatly reduces the impact of high-order faults on the resilience index. According to the IISE, the formula can be rewritten

$$R_{sys} = \sum_{w=1}^{W} P_w \sum_{j=1}^{J} \sum_{s \in \Omega_j} \left( \prod_{i \in s} p_{w,i} \right) \Delta I_{w,s} \qquad (27)$$

where $\Omega$ represents the order of the fault set, $s$ represents the number of faulty corridors, $J$ represents the highest failure order by the IISE method, $p_{w,i}$ represents the failure rate of corridor $i$ against typhoon $w$, $\Delta I_{w,s}$ represents the failure increment of the influence of state $s$ against typhoon $w$.

The physical meaning of the system-level resilience index: it is the expected load reduction of the power system under the typhoon probability model. Therefore, the smaller $R_{sys}$, the stronger the resilience of the system against typhoon disasters.

### 3.3 The corridor-level resilience index

In this paper, the corridor-level resilience index is used to analyze the fragility of the corridor. Transmission corridor $m$ resilience index $R_m$ in the power system is given by

$$R_m = R_{sys} - R_{sys.m} \qquad (28)$$

where $R_{sys.m}$ represents the system-level resilience index value when no fault occurs in transmission corridor $m$ against any typhoon disasters.

The physical meaning of the corridor-level resilience index: the drop value of the system-level resilience index caused by the failure rate of corridor $m$ becoming 0. This index shows the increase of the system-level resilience index caused by faulty corridor $m$. The larger value of $R_m$ reflects that faulty corridor $m$ causes more serious consequences to the system than other faulty corridors, so it is more necessary to strengthen corridor $m$ for the overall recovery ability of the system [56, 57].

### 3.4 Strategies for improving resilience

When the system-level resilience index $R_{sys}$ is higher than the standard value set $R_{set}$ of the system, the system resilience does not meet the requirements, and it is necessary to adopt a resilience improvement strategy, that is, to strengthen the corresponding transmission corridors. The strengthening of transmission corridors usually can be divided into the enhancement of system components and additional redundancies. The former is to replace the transmission tower or transmission line with stronger disaster resistance, for example, the traditional corner tower is replaced with the cat-head type tower with better performance in the project. The latter is to add spare transmission corridors in high failure rate areas that can reduce the failure rate of the transmission system. At the same time, the length of different transmission corridors is different and the geographical environment is different, so the cost of strengthening the transmission corridor is different. Therefore, the resilience strategy can be formulated according to the reinforcement cost, the improvement effect, and the construction difficulty, and the benefits of the strategies can be compared to determine the best reinforcement strategy.

## 3.5 Resilience assessment and improvement process

Fig. 8 shows the process of the planning-targeted resilience assessment method that considers the impact of multiple factors. According to Fig. 8, the resilience assessment and improvement process are described:

Step 1: This paper firstly analyzes the key parameters of the typhoon to generate the probability of potential typhoon occurrence.

Step 2: According to the disaster information, the comprehensive failure probability model of the system is constructed.

Step 3: According to the constraints of power flow constraints, MATPOWER software is used to calculate the optimal load reduction against various disaster scenarios.

Step 4: This paper adopts the IISE method to construct the system-level resilience index. If the index value $R_{sys}$ meets the set index value $R_{set}$, the optimal planning scheme of the system is output, otherwise step 5 is performed.

Step 5: By analyzing the corridor-level resilience indicators, this paper seeks out the weaknesses of the system and formulates strategies for improving its resilience of the system.

Step 6: According to the cost-effectiveness ratio and the construction difficulty, the optimal resilience improvement strategy is determined.

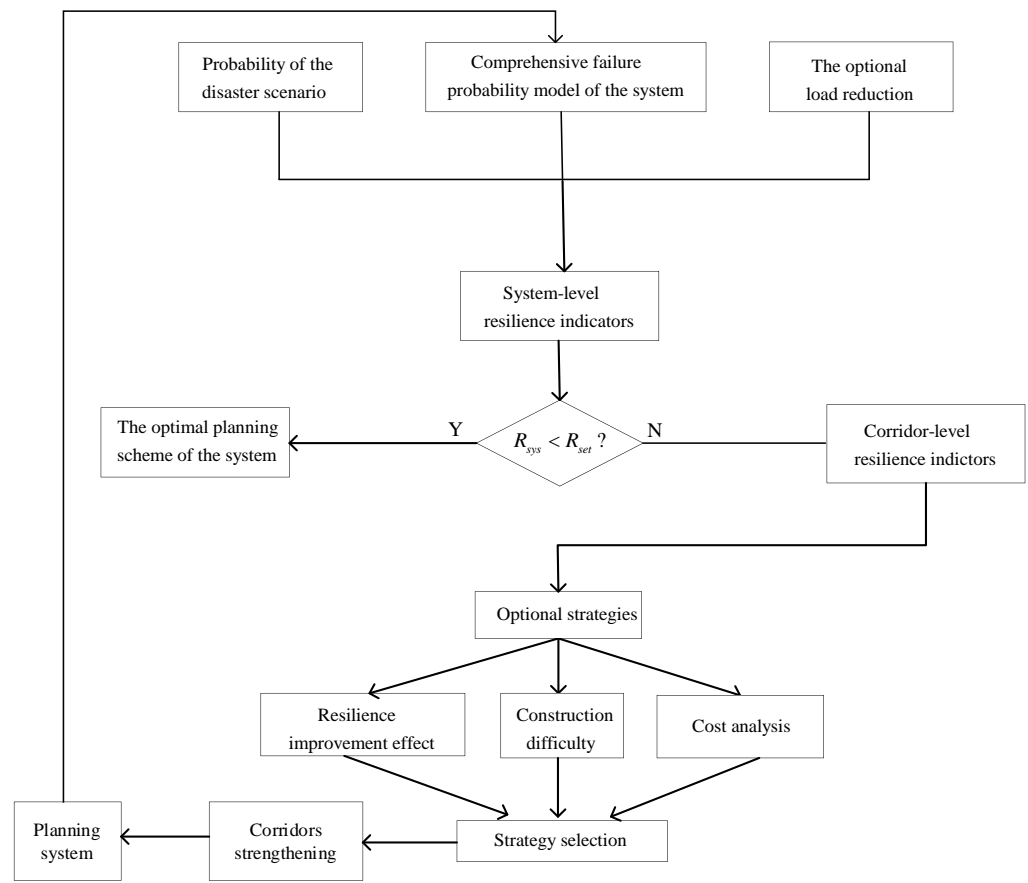

Fig. 8 The framework of resilience assessment and improvement

## 4. Case study

The IEEE RTS-79 test system is used for the verification of the effectiveness and practicability of the proposed method. The system includes 32 transmission corridors with a total transmission load level of 2850 MW [58]. At the same time, since the coast of Guangdong Province is often attacked by typhoons, we attach the test system to the coastal areas of Guangdong Province for the convenience of analysis. The meteorological data is analyzed by taking the actual data of typhoon "Mangkhut". A transmission tower is set up every 500 m, and the length of the transmission line between two adjacent transmission towers in the same corridor is 500 m. ( Each tower-line analysis unit information is actual information with approximate geographic location) At the same time, this paper adds a control test to solve the failure rate of transmission corridor based on model-driven considering multi-factor information to highlight the effectiveness of the method proposed in this paper [60]-[62]. The positions of all transmission corridors are set [6]. Fig. 9 shows the wiring location of IEEE RTS-79 and the topographic map along the coast in Guangdong province, China.

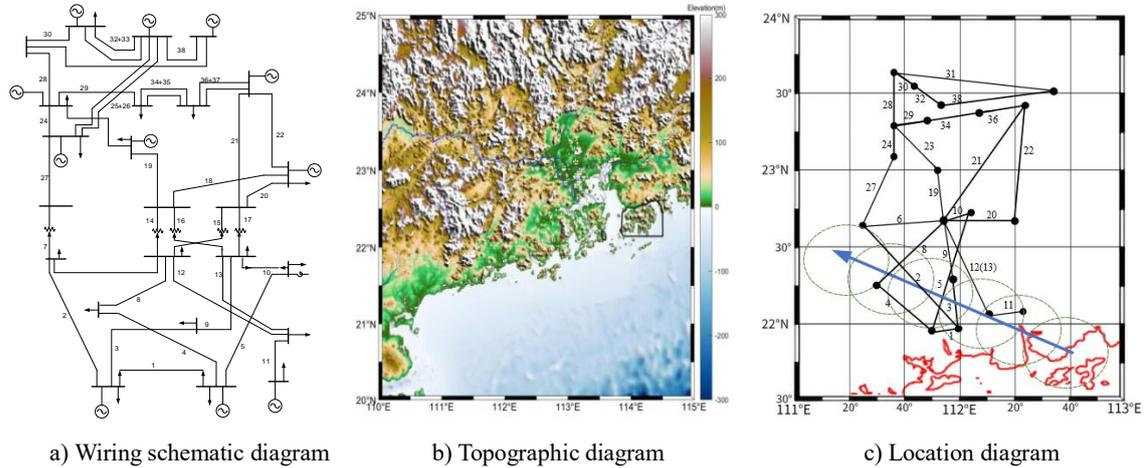

a) Wiring schematic diagram     b) Topographic diagram     c) Location diagram

Fig. 9 Wiring principle and topographic

The landing point of typhoon "Mangkhut" is 21.8°N/112.7°E, the initial pressure difference is 58hpa, and the average moving speed is 30 km/h. As shown in Fig.9 (c), the blue arrow represents the movement path of the typhoon, the green part represents the wind speed radius of the typhoon, and the red curve part represents the coastline in Guangdong Province [9]. For ease of analysis, transmission corridor 3, transmission corridor 4, transmission corridor 8, corridor 11, and transmission corridor 27 are selected as representative test corridors. Fig.10 shows the wind speed at the center of the above five transmission corridors.

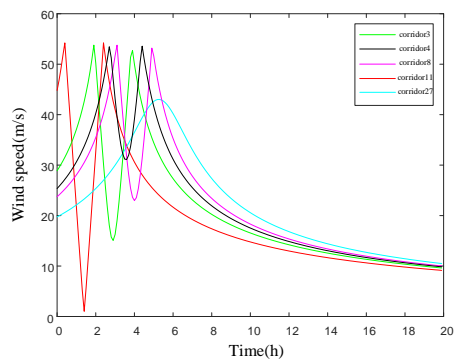

Fig. 10 Wind speed at the center of the transmission corridors

Take transmission corridor 3 as an example. When the distance between transmission corridor 3 and the typhoon center is greater than the wind circle radius, the wind speed experienced by corridor 3 increases as the distance between the typhoon center and corridor 3 decreases. When the distance is less than the wind circle radius, the wind speed in corridor 3 starts to decrease. The wind speed increases again when the wind speed center moves away from corridor 3. When the distance exceeds the wind circle radius, the wind speed gradually decreases until it reaches 0.

### 4.1 Model-driven failure rate calculation

We carry out the physical modeling analysis of the example IEEE RTS-79 test system, and calculate the failure rate of transmission corridors in each grid. The area of 1km × 1km is used to represent the transmission corridors of the same voltage level in the grid to reduce computational costs. As shown in Fig.11(a), the cumulative failure rate of the transmission corridor increases with time, and the greater the wind speed, the more severe the growth rate of the cumulative failure rate of the corridor. The cumulative failure rate of other corridors can also be calculated in the same way.

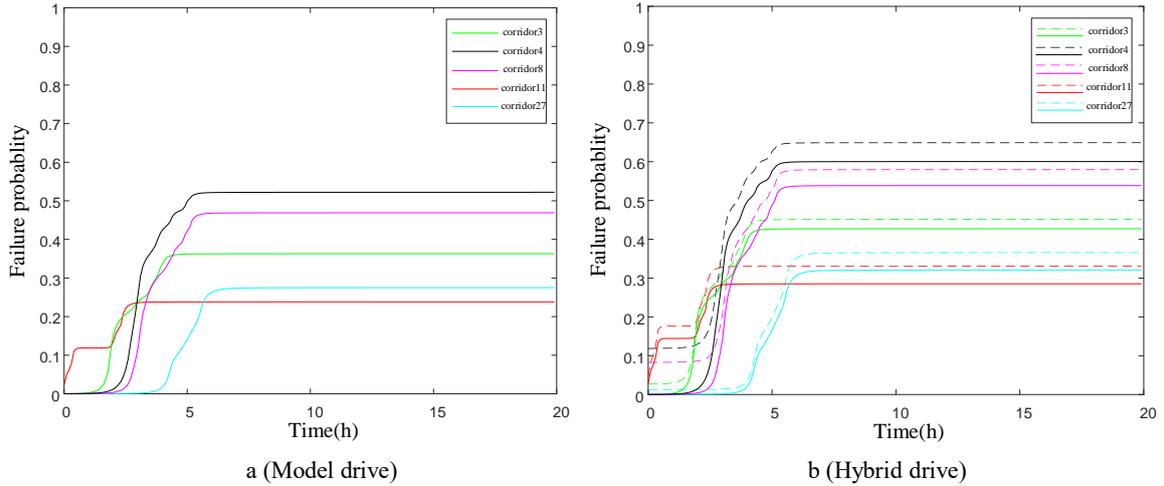

a (Model drive)    b (Hybrid drive)

Fig. 11 Failure rates of the transmission corridors

### 4.2 Data-driven failure rate calculation

This paper collects and organizes 640 sets of sample data of transmission corridors in Guangdong against typhoon disasters, and the sample information is balanced. The sample information contains seven main feature factors: maximum wind speed, rainfall intensity, altitude, slope, wind angle, design speed, and operation time. Among them, since the rainfall intensity reported by China Meteorological Administration is often the total rainfall of 24 hours, it is difficult to directly use the response analysis of the transmission corridor. Ref. [59] proposed a set of equivalent conversion methods for the rainfall amounts in 24 hours into the rainfall intensity in 10 minutes

$$R_{10\min} = 27.08 R_{24h}^{0.6021} \tag{29}$$

where $R_{24h}$ is the rainfall amounts in 24 hours, $R_{10\min}$ is the rainfall intensity in 10 minutes.

Based on the collected historical data, this paper obtains schemes 1-3 in turn by adopting the Gini index method, the OOB error method, and the entropy weight method to calculate the importance weights of each variable, as shown in Table 2.

Table 2 Feature variable importance weight evaluation results

| Schemes | Weights | | | | | | |
|---|---|---|---|---|---|---|---|
| | Max wind speed | Rainfall intensity | Altitude | Slope | Wind angle | Design wind | Operation time |
| Scheme 1 (Gini index method) | 0.236 | 0.146 | 0.182 | 0.078 | 0.105 | 0.081 | 0.169 |
| Scheme 2 (OOB error method) | 0.208 | 0.214 | 0.104 | 0.099 | 0.166 | 0.119 | 0.095 |
| Scheme 3 (Entropy weight method) | 0.158 | 0.139 | 0.186 | 0.130 | 0.129 | 0.136 | 0.122 |

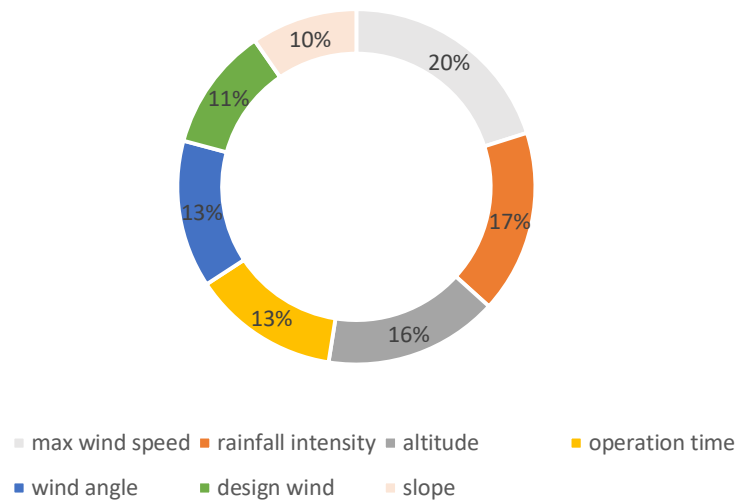

Fig. 12 Proportional distribution of different features

In Scheme 1, the maximum wind speed accounts for the highest weight ratio of 23.6%, and the slope accounts for the lower weight ratio of 7.8%. In Scheme 2, the highest proportion of the rainfall intensity is 21.4%, and the lowest proportion of the operation time is 9.5%. In Scheme 3, the altitude has the highest weight ratio of 18.6%, and the slope has a low weight ratio of 12.2% (Table 2). Due to the different evaluation principles of the three methods, the results of three methods are different. In this paper, the AHP-WAA is used for data-model screening. Fig.12 shows the average proportion distribution of each feature factor by the three schemes. Based on the average proportion of each feature factor and the evaluation of experts, the ranking of the influence of each factor on the transmission corridor is obtained: maximum wind speed > rainfall intensity > altitude >operation time= wind angle> design wind> slope. Referring to the Santy scale method, the relative importance of feature variables is obtained. The comparison of levels of different feature variables is shown in Table 3.

Table 3 Comparison of levels of different feature variables

| Comparison | Maximum wind speed | Rainfall intensity | Altitude | Operation time | Wind angle | Design wind | Slope |
|---|---|---|---|---|---|---|---|
| Maximum wind speed | 1 | 3 | 4 | 9 | 5 | 7 | 5 |
| Rainfall intensity | 1/3 | 1 | 2 | 6 | 3 | 5 | 3 |
| Altitude | 1/4 | 1/2 | 1 | 5 | 2 | 4 | 2 |
| Operation time | 1/9 | 1/6 | 1/5 | 1 | 1/4 | 1/2 | 1/4 |
| Wind angle | 1/5 | 1/3 | 1/2 | 4 | 1 | 3 | 1 |
| Design wind | 1/7 | 1/5 | 1/4 | 2 | 1/3 | 1 | 1/3 |
| Slope | 1/5 | 1/3 | 1/2 | 4 | 1 | 3 | 1 |

Then, the decision scores of each scheme are calculated according to the AHP-WAA decision-making method, and the results are shown in Table 4.

Table 4 Scheme scores

| Schemes | Scheme 1 | Scheme 2 | Scheme 3 |
|---|---|---|---|
| Weights | 0.1822 | 0.1753 | 0.1511 |

In Table 4, Scheme 1 has the highest score in scheme scores, and its evaluation results have a higher correlation to the human subjective experience. It can be learned from Table 2 that factors such as the maximum wind speed, the altitude, and the operation time have a greater impact on the failure probability, while the slope has the least impact on the failure probability. The dotted line in Fig. 11(b) represents the failure rate of the above corridors based on the model-driven consideration of the rainfall intensity, altitude, slope, wind angle, and other feature factors [60]-[62]. The solid line part of Fig. 11(b) stands for the comprehensive failure rate of the corridor under the hybrid-driven model proposed in this paper. Comparing Fig. 11(a), we can see that the failure rate of the corridor under the method proposed in this paper has a better fit with the control test. At the same time, it can be observed that the cumulative failure rate of the corridor under the hybrid-driven model is higher than that of the model drive, and the failure risk is higher than that before the correction. This is because the model drive controls other factors to ideal values, but fails to consider the influence of multiple factors on corridors.

### 4.3 Analysis of resilience index

We adopt the IISE to solve $R_{sys}$ against the typhoon disaster. For illustration purposes, the default order of enumerating failure is set to 2. Since the transmission system along the coast of Guangdong is often disturbed by typhoons, the system-level resilience index fails to reveal the corridor weaknesses. Therefore, this paper introduces the corridor-level resilience index to analyze the vulnerability of the corridor. During the calculation of the system-level resilience index, we compute the probability of typhoon occurrence, the comprehensive failure rate of the corridor, and the impact increment of each fault scenario, then we obtain the resilience index of each corridor according to equation (28). The system-level resilience indexes and the corresponding corridor-level indexes of the five most vulnerable corridors are calculated before and after the correction respectively, and the results are shown in Table 5.

Table 5 System resilience index comparison

| Model drive | | Hybrid drive | |
|---|---|---|---|
| Resilience indicator | Result (MW) | Resilience indicator | Result (MW) |
| $R_{sys}$ | $19.98 \times 10^{-5}$ | $R_{sys}$ | $23.67 \times 10^{-5}$ |
| $R_{27}$ | $8.22 \times 10^{-5}$ | $R_{27}$ | $9.19 \times 10^{-5}$ |
| $R_{11}$ | $6.03 \times 10^{-5}$ | $R_{11}$ | $6.57 \times 10^{-5}$ |
| $R_{13}$ | $3.07 \times 10^{-5}$ | $R_{4}$ | $4.07 \times 10^{-5}$ |
| $R_{12}$ | $2.94 \times 10^{-5}$ | $R_{8}$ | $4.06 \times 10^{-5}$ |
| $R_{4}$ | $2.73 \times 10^{-5}$ | $R_{13}$ | $3.71 \times 10^{-5}$ |

According to Table 5, it can be seen that the resilience index values of each transmission corridor are similar. This is because the resilience indicator is calculated taking into account the probability of typhoon occurrence in the corresponding scenario, so the overall resilience index of the corridor is relatively small, which also causes the corresponding difference between them to be small. At the same time, we can see that the $R_{sys}$ of the proposed hybrid driven approach is higher than that of the model-driven method. Due to the introduction of the multi-factor information, the cumulative failure rate of each transmission corridor of the proposed approach is relatively higher than that of the traditional model-driven method without consideration of multi-factor information. In addition, it can be found that the priority order of the transmission corridor has changed when considering the correction coefficient, corridor 4 and corridor 8 have higher priority than corridor 13. The reason for this is that the corridor-level resilience index value is closely related to its corridor failure rate. From Figure 9(b), it can be seen that the altitudes of the geographical locations of corridor 4 and 8 are significantly higher than in corridor 13. In addition, according to the National Meteorological Centre's report on Sep. 16, 2018, it showed that corridor 4 was in the area of the extraordinary rainstorm, corridor 8 was in the large rainstorm area, and corridor 13 was in the heavy rain area. In summary, the environment where corridors 4 and 8 are located is worse than that of corridor 13 without considering the influence of typhoon wind speed. Combined with the analysis of multi-factor information, the corridor-level resilience index values of corridor 4 and corridor 8 are significantly improved compared with corridor 13 after the correction, so the priority of the corridor has changed.

This paper considers the addition of the transmission corridor redundancy as the strengthening measure for the transmission system, and the strengthening cost of different transmission corridors is significantly different. Note that this work doesn't consider the construction difficulty. According to [53, 54], this paper sets the cost of the transmission corridor at 1 million $/km. In addition, planners need to comprehensively weigh the cost-effectiveness ratio of corridor enhancement strategies. When $R_{set}$ is set to $17 \times 10^{-5}$ MW, this paper proposes 6 strengthening strategies:

Strategy 1: Strengthen the redundancy of the most fragile transmission corridor 27.
Strategy 2: Strengthen redundant transmission corridors 8 and 11.
Strategy 3: Strengthen redundant transmission corridors 4 and 11.
Strategy 4: Strengthen redundant transmission corridors 3, 8 and 13.
Strategy 5: Strengthen redundant transmission corridors 3, 10 and 13.
Strategy 6: Strengthen redundant transmission corridors 8, 10 and 13.

Table 6 shows the effect of the resilience improvement and the cost of each strategy, C is the strengthening cost, RE is the reduction of the system-level resilience index, ΔRE is the percentage of reduction, and C/ΔRE is the cost-effectiveness ratio that the cost required to reduce the system-resilience index by 1%.

Table 6 Strengthening strategies

| Priority | Strategies | C ($) | RE (MW) | ΔRE | C/ΔRE ($) |
|---|---|---|---|---|---|
| 1 | Strategy 1 | $579.63 \times 10^5$ | $9.19 \times 10^{-5}$ | 38.83% | $14.93 \times 10^5$ |
| 2 | Strategy 2 | $692.35 \times 10^5$ | $10.63 \times 10^{-5}$ | 44.92% | $15.41 \times 10^5$ |
| 3 | Strategy 3 | $788.92 \times 10^5$ | $10.64 \times 10^{-5}$ | 45.11% | $17.49 \times 10^5$ |
| 4 | Strategy 5 | $1304.11 \times 10^5$ | $9.11 \times 10^{-5}$ | 38.52% | $33.86 \times 10^5$ |
| 5 | Strategy 4 | $1481.20 \times 10^5$ | $9.77 \times 10^{-5}$ | 41.30% | $35.86 \times 10^5$ |
| 6 | Strategy 6 | $1384.62 \times 10^5$ | $7.06 \times 10^{-5}$ | 30.02% | $46.12 \times 10^5$ |

As shown in Table 6, the resilience improvement effect of strategy 1 is the best and the lowest cost-effectiveness ratio is $\$14.93 \times 10^5$, and the resilience improvement effect of strategy 6 is the worst, but the cost-effectiveness ratio is $\$46.12 \times 10^5$ which is much higher than that of the other options. At the same time, if the construction of strategy 1 is difficult, strategy 2 and strategy 3 can also be considered. Both strategies can specify the system-resilience index value within $17 \times 10^{-5}$ MW, but the cost-effectiveness ratio increases slightly. Based on the above analysis, it can be concluded that strategy 1 is the best strategy.

## 5. Conclusions

At present, due to the uncertainty of the typhoon disaster information, the transmission corridor information, and the micro-topographic information, it is difficult to accurately assess the resilience of the transmission system. In addition, it is a pressing issue to be solved to deal with the difficulty of modeling the weak factors of transmission corridors caused by typhoon disasters. Therefore, we propose a planning-targeted resilience assessment framework that considers the impact of multiple factors against typhoon disasters.

1) This paper adopts the data-model hybrid-driven resilience assessment framework that fully considers the impact of various uncertain factors on system assessment against typhoon disasters, and evaluates the resilience of the transmission system in an all-around and multi-element manner. In addition, the hybrid-driven model used in this paper successfully solves the problems of difficult modeling of weak influencing factors such as the micro-topographic, and the transmission corridor information.

2) In terms of solving method, the AHP-WAA scheme-making method is proposed to determine the most accurate weight evaluation for different data-driven schemes. This method seeks the optimal data-driven evaluation result by combining subjective scoring and objective optimization, and solves the problem of large errors in evaluation results caused by different data-driven models.

3) The IEEE RTS-79 system is used to verify the performance of the method in this paper. The experimental results show that the resilience improvement index of the hybrid-driven model is more significant and adaptable than the driven model. It can accurately reveal the weaknesses of the transmission system and provide decision-makers with a more suitable resilience improvement strategy. However, the damage caused by the secondary disaster of typhoons, such as mudslides and floods, to the transmission corridor cannot be ignored in actual statistics. It is still a challenge to quantify the hazard of secondary disasters and other feature factors to the system. At the same time, the method proposed in this paper is easily affected by the feature data of the sample, and the feature factors that are not optimized and screened will cause certain interference to the accuracy of machine learning. The construction difficulty of different terrains has not been investigated yet, and the flexibility improvement strategies still need to be enriched.

Future work will focus on the optimal selection of the feature factors and the selection of the sample data. By providing the more optimized and comprehensive feature sample data, the key feature factors can be effectively screened, the prediction accuracy can be improved, and the resilience of the transmission system can be better improved. In addition, the actual transmission system will be selected in this paper to better verify the accuracy of the proposed method when the experimental conditions are satisfied. Another interesting topic is to extend the presented resilience assessment approach to analyze the resilience of integrated energy systems [63]-[65]. Simultaneously, we will also focus on the impact of terrain on the difficulty of construction in order to provide decision-makers with a more comprehensive strategy for improving resilience in future work.